# TEST OF OPTICAL STOCHASTIC COOLING IN THE IOTA RING*

V. A. Lebedev[#] and Yu. Tokpanov, FNAL, Batavia, IL 60510, USA
M. S. Zolotorev, LBNL, Berkeley, CA 94720, USA

*Abstract*

A new 150 MeV electron storage ring is being built at Fermilab. The construction of a new machine pursues two goals a test of highly non-linear integrable optics and a test of optical stochastic cooling. This paper discusses details of OSC arrangements, choice of major parameters of the cooling scheme and incoming experimental tests of the optical amplifier prototype which uses highly doped Ti-sapphire crystal as amplification medium.

## INTRODUCTION

The stochastic cooling has been successfully used in a number of machines for particle cooling and accumulation. Cooling rates of few hours required for luminosity control in hadron colliders cannot be achieved in microwave frequency range (~$10^9$-$10^{10}$ Hz) usually used in stochastic cooling. Large longitudinal particle density, $N/\sigma_s$, used in such colliders requires an increase of cooling bandwidth by a few orders of magnitude. To achieve such increase one needs to make a transition to much higher frequencies. A practical scheme operating in the optical frequency range was suggested in Ref. [1]. The method is named the optical stochastic cooling (OSC). It is based on the same principles as the stochastic cooling but uses much higher frequencies. Thus it is expected to have a bandwidth of ~$10^{14}$ Hz and can create a way to attain required damping rates. Fermilab plans to make an experimental test of the OSC in IOTA ring [2].

In the OSC a particle emits e.-m. radiation in the first (pickup) wiggler. Then, the radiation amplified in an optical amplifier (OA) makes a longitudinal kick to the same particle in the second (kicker) wiggler as shown in Figure 1. A magnetic chicane is used to make space for the OA and to delay a particle so that to compensate for a delay of its radiation in the OA resulting in simultaneous arrival of the particle and its amplified radiation to the kicker wiggler. A particle passage through the chicane has a coordinate-dependent correction of particle longitudinal position which, consequently, results in a correction of relative particle momentum, $\delta p/p$, with amplitude $\xi_0$ so that:

$$\delta p / p = -\xi_0 \sin(k \Delta s) . \quad (1)$$

Here $k = 2\pi/\lambda$ is the radiation wave-number,

$$\Delta s = M_{51} x + M_{52} \theta_x + M_{56} (\Delta p / p) , \quad (2)$$

is the particle displacement on the way from pickup to kicker relative to the reference particle which experiences zero displacement and obtains zero kick, $M_{5n}$ are the elements of 6x6 transfer matrix from pickup to kicker, $x$, $\theta_x$ and $\Delta p/p$ are the particle coordinate, angle and relative momentum deviation in the pickup.

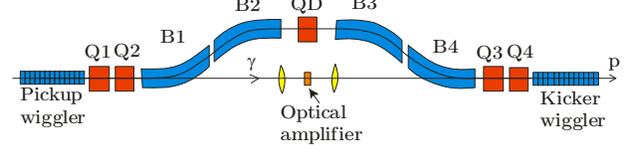

Figure 1: OSC schematic.

For small amplitude oscillations the horizontal and vertical cooling rates per turn are [4]:

$$\begin{bmatrix} \lambda_x \\ \lambda_s \end{bmatrix} = \frac{k\xi_0}{2} \begin{bmatrix} M_{56} - \tilde{M}_{56} \\ \tilde{M}_{56} \end{bmatrix} , \quad (3)$$

where $\tilde{M}_{56} = M_{51} D_p + M_{52} D'_p + M_{56}$ is the partial slip-factor introduced so that for a particle without betatron oscillations and with momentum deviation $\Delta p/p$ the longitudinal displacement relative to the reference particle on the way from pickup to kicker is equal to $\tilde{M}_{56} \Delta p / p$, and $D$ and $D'$ are the dispersion and its derivative in the pickup. We assume that there is no x-y coupling in the chicane. Introduction of x-y coupling outside the cooling area allows a redistribution of the horizontal damping rate between two transverse planes. The sum of damping rates is: $\Sigma \lambda_n = k \xi_0 M_{56}/2$.

A non-linear dependence of kick on $\Delta s$ in Eq. (1) results in a dependence of damping rates on amplitudes [3]:

$$\lambda_x(a_x, a_s) = (2 J_0(a_s) J_1(a_x) / a_x) \lambda_x , \\ \lambda_s(a_x, a_s) = (2 J_0(a_x) J_1(a_s) / a_s) \lambda_s , \quad (4)$$

where $a_x$ and $a_s$ are the amplitudes of longitudinal particle motion due to betatron and synchrotron oscillations expressed in the units of e.-m. wave phase:

$$a_x = k \sqrt{\varepsilon_1 \left( \beta_p M_{51}^2 - 2\alpha_p M_{51} M_{52} + (1 + \alpha_p^2) M_{52}^2 \right)}, \\ a_p = k |\tilde{M}_{51}| (\Delta p / p) , \quad (5)$$

$\varepsilon_1$ is the Courant-Snyder invariant of a particle, and ($\Delta p/p$) is the amplitude of particle synchrotron motion. As one can see from Eq. (4) each damping rate changes its sign if any of amplitudes exceeds the first root, $\mu_0$, of the Bessel function $J_0(x)$. That determines that at the cooling area boundary $a_x = \mu_0$ or $a_y = \mu_0$ where $\mu_0 \approx 2.405$. That yields the cooling area boundaries:

$$\varepsilon_{max} = \frac{\mu_0^2}{k^2 \left( \beta_p M_{51}^2 - 2\alpha_p M_{51} M_{52} + (1 + \alpha_p^2) M_{52}^2 \right)} , \quad (6)$$

$$(\Delta p / p)_{max} = \mu_0 / (k |\tilde{M}_{51}|) .$$

Although $M_{56}$ and, consequently, the sum of damping rates depend only on focusing inside the chicane, $\tilde{M}_{56}$ and the ratio of damping rates depend on the dispersion at the chicane beginning, i.e. on the ring dispersion. Eq. (3) yields the ratio of damping rates,

___________________
*Work supported by Fermi Research Alliance, LLC, under Contract No. DE-AC02-07CH11359 with the United States Dep. of Energy
#val@fnal.gov

$$\lambda_x / \lambda_s = M_{56} / \tilde{M}_{56} - 1 . \quad (7)$$

The relative cooling ranges are introduced as ratios of cooling area boundaries $(\Delta p/p)_{max}$ and $\varepsilon_{max}$ to the rms values of momentum spread, $\sigma_p$, and horizontal emittance, $\varepsilon$, established by equilibrium between SR damping and heating. That yields:

$$n_{\sigma s} = (\Delta p / p)_{max} / \sigma_p , \quad n_{\sigma x} = \sqrt{\varepsilon_{max} / \varepsilon} . \quad (8)$$

As one can see the transverse cooling range does not depend on the dispersion but depends on the beta-function in the pickup wiggler.

## BEAM OPTICS

The OSC system will take one of four straight sections with length of ~4 m. The cooling chicane consists of four dipoles with parallel edges (see Figure 1.) To obtain transverse cooling $M_{56}$ and $\tilde{M}_{56}$ have to be different. It is achieved by placing a defocusing quad in the chicane center. Leaving only leading terms in the thin lens approximation one obtains:

$$M_{56} \approx 2\Delta s, \quad \tilde{M}_{56} \approx 2\Delta s - \Phi D^* h ,$$
$$\lambda_x / \lambda_s \approx \Phi D^* h / (2\Delta s - \Phi D^* h). \quad (9)$$

Here $\Delta s$ and $h$ are the path lengthening and the trajectory offset in the chicane, $\Phi=1/F$ is the defocusing strength of the quad, $D^*$ is the dispersion in the chicane center, and we assume that $dD/ds=0$. The latter condition is required to minimize the equilibrium beam emittance. Similarly, using Eq. (8) one obtains estimates for the cooling ranges

$$n_{\sigma p} \approx \mu_0 / \left((2\Delta s - \Phi D^* h) k \sigma_p\right),$$
$$n_{\sigma x} \approx \mu_0 / \left(2kh\Phi\sqrt{\varepsilon\beta^*}\right), \quad (10)$$

where $\beta^* \approx L^2 / \beta$ is the beta-function in the chicane center, and $L$ is its half-length. As one can see from Eqs. (9) and (10) $\Phi D^* h$ determines the cooling dynamics. Using the bottom equation in (10) we obtain:

$$\Phi D^* h \approx \frac{\mu_0}{2kn_{\sigma x}} \sqrt{\frac{A^*}{\varepsilon}} , \quad (11)$$

where $A^* = D^{*2}/\beta^*$ is the dispersion invariant in the chicane center. As one can see the given wave-length, beam emittance and transverse cooling range uniquely determine $\Phi D^* h$ and, consequently, the cooling dynamics.

The choice of the wavelength of 800 nm was determined by choice of OA. Although the delay in the OA of 2 mm is quite small it still presents considerable challenges to the beam optics. Without redistribution of decrements it yields an acceptably low longitudinal cooling range, $n_{\sigma p}$ <1. Therefore a redistribution of decrements is required but its value is limited by reduction of the transverse cooling range. Decreasing the ring energy from 150 to 100 MeV reduces the equilibrium emittance and momentum spread and allows obtaining sufficiently large cooling ranges of >2 for both planes. Operation at the coupling resonance redistributes SR cooling rates between horizontal and vertical planes resulting approximately equal equilibrium transverse emittances and, consequently, twice smaller horizontal emittance. Tables 1 and 2 present the main parameters of the ring. The ring optics is symmetric relative to the chicane center. Figure 2 presents the dispersion and beta-functions for half of the ring. The optics was designed to minimize the equilibrium emittance. It requires a minimization of the dispersion invariant $A = (1+\alpha^2)D^2/\beta + 2\alpha DD' + \beta D'^2$ in the dipoles. Its initial value of 11 m determined by requirements of OSC is quite large and, if not addressed, would result in an unacceptably large emittance. An adjustment of horizontal betatron phase advances between dipoles was crucial to obtain tolerable transverse emittances. Figure 3 presents the value of $A$ along half of the ring.

Table 1: Main Parameters of IOTA storage ring for OSC

| Circumference | 40 m |
|---|---|
| Nominal beam energy | 100 MeV |
| Bending field | 4.8 kG |
| Tunes, $Q_x/Q_y$ | 6.36/2.36 |
| Number of particles | $5 \cdot 10^8$ |
| Transverse emittances, $\varepsilon = \varepsilon_x = \varepsilon_y$, rms | 11.5 nm |
| Rms momentum spread, $\sigma_p$ | $1.23 \cdot 10^{-4}$ |
| SR damping times (ampl.), $\tau_s / (\tau_x = \tau_y)$ | 1.4 / 0.67 s |

Table 2: Major parameters of cooling chicane

| Delay in the chicane, $\Delta s$ | 2 mm |
|---|---|
| Horizontal beam offset, $h$ | 20.1 mm |
| $M_{56}$ | 3.95 mm |
| Dispersion in the chicane center, $D^*$ | 307 mm |
| Beta-function in the chicane center, $\beta^*$ | 8.59 mm |
| Cooling rates ratio, $(\lambda_x = \lambda_y)/\lambda_s$ | 1.18 |
| Cooling ranges (before OSC), $n_{\sigma x}/n_{\sigma s}$ | 2.1 / 3.2 |
| Dipole magnetic field | 4.22 kG |
| Dipole length | 10 cm |
| Strength of central quad, $GdL$ | 1.58 kG |

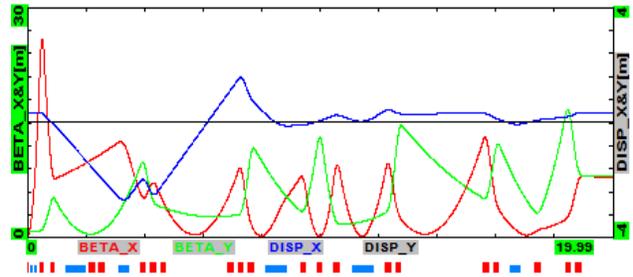

Figure 2: Optics functions for IOTA half ring starting from the OSC section.

## OPTICAL AMPLIFIER

Choice of Ti:Sapphire OA has a few advantages. First, it has quite wide bandwidth (~20% FWHM). Second, it allows operation in the CW regime; and third, it can deliver significant amplification with only ~1 mm signal delay. For an OA prototype we bought a highly doped (0.5%wt $Ti_2O_3$) 2 mm thick Ti: Sapphire crystal from GT Crystal Systems. An estimated low power gain is ~100 (20 Db) with pumping power density of 1.8 MW/cm$^2$.

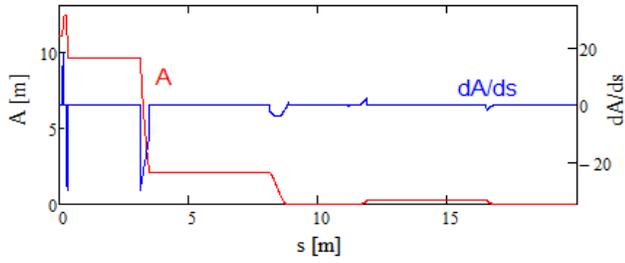

Figure 3: Dependence of the dispersion invariant, $A$, and its derivative for IOTA half-ring starting from the OSC section. $dA/ds$ is non-zero at dipoles.

Optical system has two lenses (Figure 1). The first lens focuses the beam radiation on the Ti:Sapphire crystal, and the second one focuses the amplified radiation into the kicker wiggler. The lenses have the same focal length and radius. The focal length was chosen to minimize the size of e.-m. radiation on the Ti:Sapphire crystal. A longer focal length decreases the spot size at the crystal entrance but increases the waist and effects of the field of depth. An optimal value of the focal length is 8 cm with lens radius of 3.5 mm. The lens radius was determined by a compromise between the damping rates growth with the radius and an increase of signal delay leading to a drastic deterioration of beam optics. The choice of lens parameters results in that the most radiation is within cylinder with 30 μm radius where amplification can happen. The radiation of pumping laser follows along beam radiation axis. Windows in vacuum chamber located along the axis allow the pumping light to get in and out together with the amplified radiation. The radiation of pumping laser comes through the same lenses. Additional matching lenses are installed outside the vacuum chamber. The pumping laser radiation is focused into 30 μm cylinder with close to rectangular power density distribution. It determines the laser power to be ~50 W. To reduce the effect of the depth of field a 50 cm gap between the cooling chicane and the wigglers is introduced.

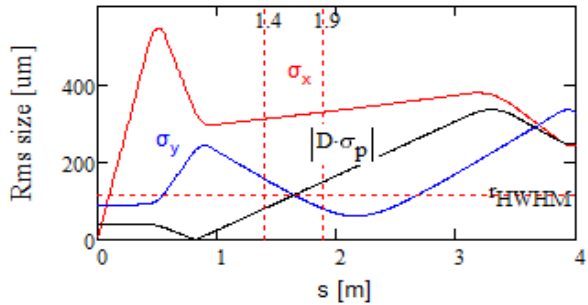

Figure 4: Rms beam sizes (horizontal – $\sigma_x$, vertical – $\sigma_y$, and due to momentum spread - $|D\sigma_p|$) in vicinity of cooling chicane starting from the center of OSC section. Vertical lines mark location of wiggler. The horizontal line shows diffraction limited spot size (HWHM) of amplified radiation in the kicker wiggler.

Cooling the OA to liquid nitrogen temperature is required. It increases the crystal thermal conductivity resulting in an acceptable temperature difference across the crystal (~8K) and thermal stresses. Cooling also reduces the dependence of refraction index on temperature and, thus, optics distortions related to the high pumping power.

Recently we assembled a prototype OA which will be operating in a pulsed regime to avoid problems with cooling required for CW regime. The goal of the measurements is to determine the dependence of phase and amplitude of amplified signal on frequency. It is achieved by inserting the OA into one leg of Michelson interferometer. The first measurements are expected soon.

## COOLING RATES

Parameters of undulators and cooling rates are shown in Table 3. The undulator period was chosen so that the wavelength of the radiation at zero angle would be at the band boundary (750 nm). The cooling rates were computed using formulas developed in Ref. [4] where additional averaging over amplifier band was applied resulting in ~5% reduction of rates. It was assumed that only radiation within 2 mrad enters the OA. It sets the aperture of optical system to 2·3.5 mm and the upper boundary of the band to 850 nm. The effects of e.-m. wave dispersion in the OA amplifier were neglected. It was assumed that the dispersion in the amplifier is included into the gain, *i.e.* $G = 10$ implies an amplitude amplification of 10. Consequently, the dispersion makes the power gain to be somewhat larger than $G^2$. We also neglected incomplete overlap of light and particle beams in the kicker undulator at the beginning of cooling process when the beam size is determined by SR. Figure 4 shows the rms beam sizes in the cooling section for the uncooled beam. One can see that the beam sizes exceed the radiation spot size. It results in that the initial cooling rates are about factor of 5 lower. For the chosen number of particles in the bunch the system operates slightly below optimal gain at the beginning of cooling process. We would like to stress that even in the absence of amplification (passive system, $G = 1$) the OSC damping exceeds SR damping by more than an order of magnitude (see Tables 1 and 3).

Table 3: Main parameters of OSC

| | |
|---|---|
| Undulator parameter, K | 0.6 |
| Undulator period | 4.92 cm |
| Radiation wavelength at zero angle | 750 nm |
| Number of periods, $m$ | 10 |
| Total undulator length, $L_w$ | 0.50 m |
| Length from OA to undulator center | 1.65 m |
| Amplifier gain (amplitude) | 10 |
| Telescope aperture, $2a$ | 7 mm |
| Lens focal length, $F$ | 80 mm |
| Damp. rates ($x=y/s$) | 200/170 s$^{-1}$ |